\documentclass[aps,twocolumn,showpacs,amsmath,amssymb,pra,superscriptaddress,floatfix,longbibliography]{revtex4-1}

\usepackage{bm}
\usepackage{physics}
\usepackage{braket}
\usepackage{amsmath}
\usepackage{amssymb}
\usepackage{mathtools}
\usepackage{graphicx}
\usepackage{dcolumn}
\usepackage{bm}
\usepackage{multirow}
\usepackage{leftidx}
\usepackage{color}
\usepackage{float}
\usepackage{amsfonts}
\usepackage{indentfirst} 
\usepackage[german,english]{babel}
\usepackage{cleveref}
\usepackage{gensymb}
\usepackage[thinc]{esdiff}
\graphicspath{}
\begin{document}


\title{The effect of orientation of Rydberg atoms on their collisional ionization cross section} 



\author{Akilesh Venkatesh}
\affiliation{Department of Physics and Astronomy, Purdue University, West Lafayette, Indiana 47907, USA}

\author{Francis Robicheaux}
\affiliation{Department of Physics and Astronomy, Purdue University, West Lafayette, Indiana 47907, USA}
\affiliation{Purdue Quantum Center, Purdue University, West Lafayette, Indiana 47907, USA}

\date{\today}

\begin{abstract}
Collisional ionization between two Rydberg atoms in relative motion is examined. A classical trajectory Monte Carlo method is used to determine the cross sections associated with Penning ionization. The dependence of the ionization cross section on the magnitude and the direction of orbital angular momentum of the electrons and the direction of the Laplace-Runge-Lenz vector of the electrons is studied. 
For a given magnitude of angular momentum, there can exist a difference of a factor of up to $\sim2.5$ in the ionization cross section between the orientation with the highest and the lowest ionization cross section.
The case of exchange ionization is examined and its dependence on the magnitude of angular momentum is studied. 
\end{abstract}


\maketitle

\section{Introduction} \label{Intro}
The study of Rydberg atoms has seen considerable progress in the last few decades~\cite{specialissue_rydbergphysics}. The highly excited state of the electrons in these Rydberg atoms give rise to many interesting properties such as controllable long range interactions~\cite{longrange_controllable_experiment}, strong response to electric and magnetic fields~\cite{electricfields,magneticfields} and classical behaviour of the valence electrons, all of which have seen considerable analysis~\cite{gallagherbook_1994}. Their manipulable interactions have enabled the study of quantum entanglement effects across multiple atoms and by extension brought about the pursuit for robust qubits built from neutral atoms~\cite{neutralatomQC_JphysB}. 

The highly excited nature of the valence electron in Rydberg atoms, makes them susceptible to ionization either due to collisions involving Rydberg atoms or  interaction with blackbody radiation~\cite{Blackbody_ionization}. In this paper, we focus exclusively on collisions between Rydberg atoms that can lead to one of the atoms becoming ionized through Penning ionization. In Penning ionization, two highly excited atoms with principal quantum number $n$, collide to give a positive ion, a free electron and the other atom with its valence electron having a principal quantum number $n'$. If there is no energy transferred to the electrons from the translational kinetic energy of the atoms, it can be shown that $n'< n/\sqrt{2}$ for the process to conserve energy. 

Rydberg atoms enjoy considerable separation when they are prepared because of Rydberg blockade effects~\cite{rydbergblockade_tong, rydbergblockade_2}. However, collisions between Rydberg atoms can still occur because of their strong interaction. These atoms interact due to van der Waals forces, dipole-dipole forces or other higher multi-pole moments depending on the distance of separation and the nature of the electronic states. Even if these atoms were initially at rest, these interactions could lead to their eventual collisional ionization ~\cite{robicheaux2005ionization}. More recently, the effect of van der Waals and dipole-dipole forces on the collisional cross section between Rydberg atoms was examined~\cite{interaction_inducedcollision} theoretically and was shown to agree reasonably with experiment.

The thermal energies of Rydberg atoms can also lead to collisional ionization. The recent work by Fields et al.~\cite{dunning_destructioncollision} not only provided an idea of the scale and the physics of destruction of Rydberg atoms from collisional ionization but also verified experimentally once again the results of a classical trajectory Monte Carlo approach for collision between two Rydberg atoms. They concluded that these collisional cross sections were significant and comparable to that of collisions between hard spheres of size comparable to the Rydberg orbit. These collisions can also be a source of transition from Rydberg atoms to ultra-cold plasma~\cite{Ryd_ultracoldplasma1,Ryd_ultracoldplasma2}. Effectively, these collisions can lead to a significant loss of the prepared Rydberg atoms and can be a cause for concern in experiments. 

Inspired by the recent results by Fields et al.~\cite{dunning_destructioncollision}, here we try to answer the next relevant question i.e. the effect of orientation of Rydberg atoms on their collisional ionization cross section. Two quantities that should adequately characterize the orientation of the colliding Rydberg atoms are the direction of angular momentum and direction of Laplace-Runge-Lenz vector. The research until this point, have been carried out either by assuming some arbitrary choice for the direction and magnitude of angular momentum and the direction of the Laplace-Runge-Lenz vector (see Figs.~\ref{Diagram_fris} and \ref{Diagram_pan}) or the results are averaged over them. There has not been, to the best of our knowledge, an analysis of their effect on the ionization cross section. In this paper we vary these parameters in a systematic way for a couple of orientations to begin understand their impact on the ionization cross section. There is a wide variety of combination of parameters that could be explored;  only a few cases were investigated to limit the size of this study. One restriction on the calculation was determined by the experimental arrangement in Ref.~\cite{dunning_destructioncollision}: we have the two atoms excited to the same state and interacting through the difference in their thermal, center of mass velocities.

The ionization cross section in general is expected to have some dependence on the magnitude of the angular momentum as it determines the eccentricity of the orbit of the electron. Our investigations of the effect of orientation reveal that for a given magnitude of angular momentum, there can exist a difference of a factor of $\sim$~2.5 in the ionization cross section between the orientation with the highest and the lowest cross sections. This can be relevant because the currently proposed methods~\cite{dunning_destructioncollision} to decrease the collisional loss of Rydberg atoms is to either lower the temperature to increase the timescale for collisional destruction or to create the Rydberg atoms in a spaced-out manner, both of which could be difficult to implement. The results in this paper can provide an additional perspective into the physics of collisional ionization with possible insights to minimize it. However, note that our results are not well-suited for application in ultra-cold Rydberg gas experiments such as Ref.~\cite{ultracoldgas_recent} as they operate in a parameter regime which is outside the scope of this paper.

Following this, we briefly consider the case of exchange ionization and study its dependence on the magnitude of angular momentum.  By exchange ionization we refer to a scenario where one of the approaching Rydberg atoms loses its electron but captures the electron from the other atom leaving the other atom effectively ionized.

One may also be interested to study the cross sections for a double ionization process in which both the colliding atoms end up being ionized, but the calculations reveal that in order for double ionization to be of significance the velocities have to be much higher and of the same order as the orbital velocity of the electrons~\cite{olson1979ionization_vcm}.

On the experimental side, Rydberg atoms are usually synthesised with the valence electron being in a state with low orbital angular momentum $l$. Over the last three decades, there have been several theoretical proposals to obtain high $l$ circular states and high $m$ states which have been followed up by successful experiments ~\cite{richardscircular,highm_Koch_1980, Huler1983_high_m, brecha1993circular_RBhot,Delande_circulartheory,delandeexperiment_circular,fastcircular,circular_theoretical}. Given the experimental progress, we believe that the manipulations proposed in this paper are within reach of current experimental techniques. 

This paper is organized as follows: in Sec. \ref{Methods}, we describe the approach to analyse collisional ionization. A discussion of the numerical method used and convergence is included. In Sec. \ref{applications}, we apply the method described in Sec. \ref{Methods} to different orientation of the two Rydberg atoms and study its effect on the ionization cross section.


Unless otherwise stated, atomic units will be used
throughout this paper.
\section{Methods and Modelling} \label{Methods}
A classical approach is used to model the pair of Rydberg atoms and their scattering. This can be justified in three ways: First, this is reasonable from the classical correspondence principle given the large principal quantum number $n$ of the Rydberg atoms. Second, we are interested in the ionization cross section with little focus on the nature of the final state of the other electron. Therefore a classical treatment of ionization is desirable as it yields a good approximation of the ionization cross section with an averaging over all possible final states ($n$,$l$,$m$) of the electrons involved. Lastly, the final state of each electron will be in energy states which are well above the ones where quantum effects are important. Under these conditions, there is strong experimental support for such a classical approach~\cite{olson_experiment,perumal2001_classical,amthor2007modeling,dunning_destructioncollision}.

Each Rydberg atom is modelled as having a nucleus with a unit charge and an electron in a classical Keplerian orbit~\cite{keplerianorbit} around the center of mass of the atom with the interaction being purely Coulombic. The initial state of the electron in each atom up to an orbital phase angle, is characterised by its energy, angular momentum and Laplace-Runge-Lenz vector. The energy chosen, corresponds to a Bohr orbit of principal quantum number $n$. The particular choice of $n$ is not important as the classical results for the collisional cross section scale with $n^4$, provided you also scale the velocity of the atoms by $1/n$. For our calculations, we choose Rubidium-85 with $n=60$. The orbital angular momentum $l$ is chosen to have a value in the range of (0, $n$]. The case of $l = n$ corresponds to the special case of Bohr orbits which are ideally circular. Decreasing $l$ below $n$ increases the eccentricity of the orbit. Therefore, the chosen range of angular momentum covers the entire range of eccentricities for the electron's orbit. The direction of the Laplace-Runge-Lenz vector is varied during the course of the calculations to study its effect on the ionization cross section. The initial position and velocity of the electron is determined up to an orbital phase angle which is randomized. 


The force exerted by particle j on particle i is given by,
\begin{equation} \label{netforce}
\boldsymbol{F_{ij}} =  \frac{q_i q_j }{ |\boldsymbol{r_i} - \boldsymbol{r_j} |^3 } (\boldsymbol{r_i} - \boldsymbol{r_j} )
\end{equation}
Here, the indices $i,j \in$ \{1,2,3,4\} and $i \neq j $. The quantity $q_i$ refers to the charge of particle $i$ and the quantity $r_i$ refers to the position vector of particle $i$. The equations of motion can then be obtained by solving the following two equations:

\begin{equation} \label{acceleration}
\diff{\boldsymbol{ v_{i}} }{t} = \frac{\sum_{j}\boldsymbol{F_{ij}}}{m_i} 
\end{equation}

\begin{equation} \label{velocity}
\diff{\boldsymbol{ r_{i} } }{t} = \boldsymbol{v_{i}}
\end{equation}
where, the terms $\boldsymbol{ v_{i}}$,  $\boldsymbol{ r_{i} }$ and $m_i$ refers to the velocity, position and mass of the i\textsuperscript{th} particle respectively.

\begin{figure}
\resizebox{80mm}{!}{\includegraphics{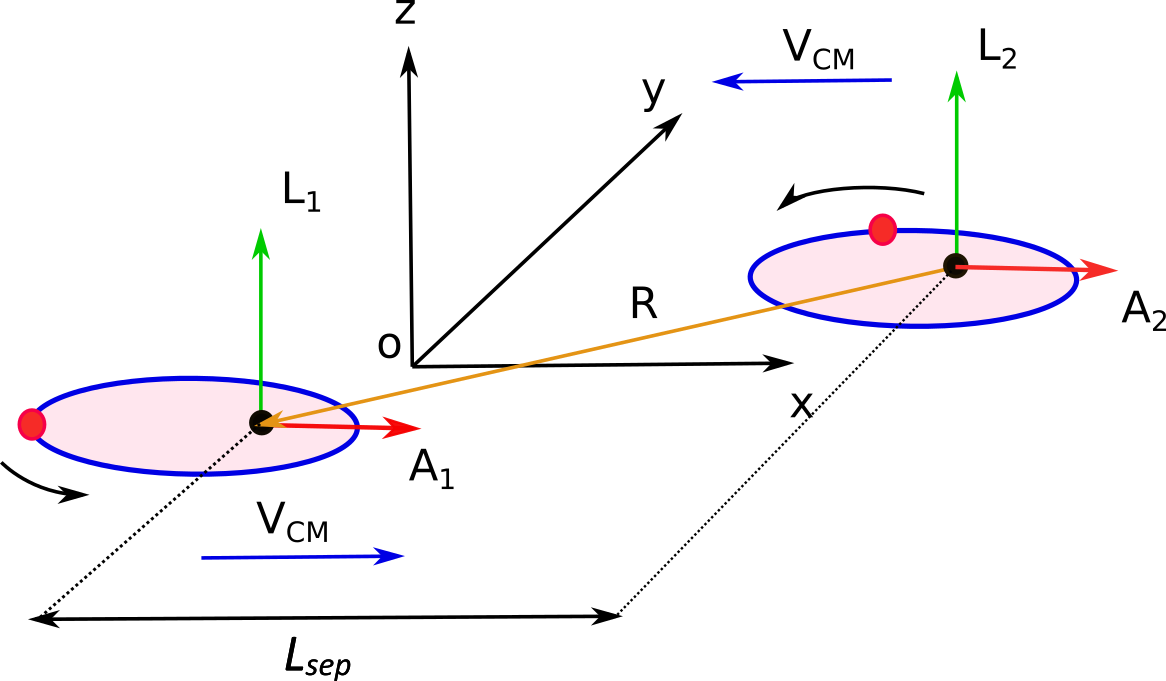}}
\caption{\label{Diagram_fris}
A schematic diagram of the two Rydberg atoms for the initial orientation of type-Frisbees. The Red dot on the edge of each ellipse represents an electron and the black dot at the focus of each ellipse represents an ion. The angular momentum vector $\boldsymbol{L}$, the Laplace-Runge-Lenz vector $\boldsymbol{A}$ and $\boldsymbol{V_{CM}}$ are properties of each atom when the separation vector $\boldsymbol{R}$  goes to infinity. To emulate recent experiments~\cite{dunning_destructioncollision}, all calculations are performed with $\boldsymbol{L_1}$ = $\boldsymbol{L_2}$, $\boldsymbol{A_1}$ = $\boldsymbol{A_2}$ and $\boldsymbol{V_{CM}}$ = $10^{-4}$ ~a.u.
Note that the orbits of the electrons are in a plane parallel to the x-y plane and $L_{sep}$ is measured along the x-axis.
}
\end{figure}

\begin{figure}
\resizebox{80mm}{!}{\includegraphics{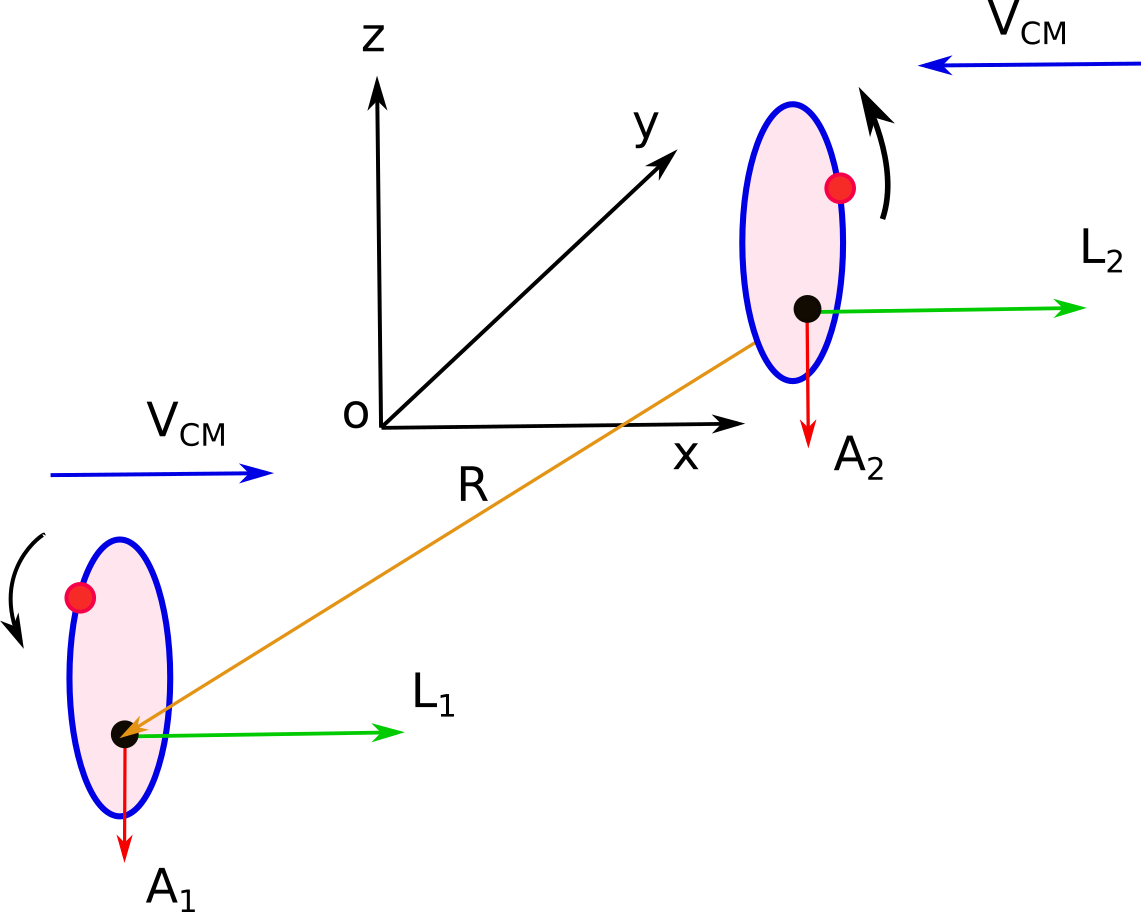}}
\caption{\label{Diagram_pan}
A schematic diagram of the two Rydberg atoms for the initial orientation of type-Cymbals. The notation is the same as in Fig.~\ref{Diagram_fris}. Note that the orbits of the electrons are in a plane parallel to the y-z plane.
}
\end{figure}

The two atoms start with a separation of $L_{sep}$, measured along the x-axis~(see Fig.~\ref{Diagram_fris}). The value for $L_{sep}$ is motivated by the minimum $L_{sep}$ required for convergence (refer Sec. \ref{convergence}) of the cross section. There still remains 2 degrees of freedom for the position of center of mass (CM) of each atom. The y and z coordinates of the center of mass of atom 1 is randomly (uniform) placed inside a disk of radius $0.5~b_{max}$ centered on the x-axis and is parallel to the y-z plane (refer Sec. \ref{convergence} for $b_{max}$). The CM of atom 2 is positioned such that the total center of mass of the two atoms lies at the origin.

At the initial time, the CM of the first atom is imparted a net velocity in the positive x-direction and the CM of the second atom in the negative x-direction. As the two Rydberg atoms drift towards each other, one of them may ionize depending on the initial conditions. In our calculations, we consider the electron to be ionized if the distance to any of the electrons from the origin of the coordinate system is greater than the initial length of separation between the two Rydberg atoms, $L_{sep}$. 

We define two quantities which serve as a convenient time scale and length scale in the problem. $T_{Ryd} = 2 \pi n^3 $ is the time period of a Bohr orbit for the chosen principal quantum number $n$ and  $R_{Ryd} = n^2$ is the radius of the same Bohr orbit.

The aim is to study the ionization cross section as a function of the direction and magnitude of the orbital angular momentum and the direction of the Laplace-Runge-Lenz vector of the electron. In order to determine the cross section, we resort to a Monte Carlo approach. We estimate the cross section by first calculating the probability of ionization from a set of 10,000 Monte Carlo runs by randomly varying the initial orbital phase angle of both the electrons and the impact parameter. The randomized initial conditions form the population of a micro-canonical statistical distribution of the phase space~\cite{abrines_perceival_1966} with an additional requirement that the orbital angular momentum of the electrons remain fixed. The initial values for the $V_{CM}$ of the atoms, energy of each electron, direction and the magnitude of the angular momentum of each electron all remain the same during each set.

The orbital phase angle of the electron in the first atom is randomized by allowing the electron to dynamically evolve with time in the absence of the second atom for a random duration between 0 - $T_{Ryd}$. For randomizing the orbital phase angle of the second electron, a small distance which randomly varies between 0 to $V_{CM} \cdot T_{Ryd}$ is added to the length of the separation, $L_{sep}$.

The cross section is typically calculated using the expression $\int 2\pi b P(b) db$ where, the probability of ionization as a function of the impact parameter ($P(b)$) is integrated over all possible impact parameters~\cite{dunning_destructioncollision}. Here using Monte Carlo
sampling described below Eq.~(\ref{velocity}), the total probability of ionization $P_{ion}$, can be calculated by determining the fraction of total runs that result in ionization and using the following expression for cross section $\sigma$: 
\begin{equation} \label{cross_sec_defn}
    \sigma =  \pi b_{max}^2 P_{ion}
\end{equation}
where, $b_{max}$ is the maximum value of the impact parameter that results in ionization. Note that it is convenient to scale the ionization cross section by $\pi(2R_{Ryd})^2$, given that  $R_{Ryd}$ is the radius of the Rydberg atom. We define the scaled cross section $\sigma_{scal}$ as,
\begin{equation} \label{scaled_cross_sec_defn}
    \sigma_{scal} =  \frac{\sigma}{\pi(2R_{Ryd})^2}
\end{equation}

We restrict our discussions to symmetric collisions in which the two colliding Rydberg atoms have the same $n$ value and both the electrons have the same orbital angular momentum $\boldsymbol{L}$ and the Laplace-Runge-Lenz vector $\boldsymbol{A}$~\cite{dunning_destructioncollision}. One may refer to Refs.~\cite{tomandjerry,nonsymmetric_penningionization} for discussions pertaining to asymmetric collisions. Here, the Laplace-Runge-Lenz vector $\boldsymbol{A}$ is defined as, 
\begin{equation}
    \boldsymbol{A} = \boldsymbol{p} \times \boldsymbol{L} - \boldsymbol{\hat{r}}
\end{equation}
where $\boldsymbol{p}$ and $\boldsymbol{L}$ indicates the linear momentum and the angular momentum of the electron with respect to the nucleus. The quantity $\boldsymbol{\hat{r}}$ indicates the unit position vector of the electron measured with nucleus as the origin. The magnitude of the Laplace-Runge-Lenz vector is proportional to the eccentricity of the orbit. It is to be noted that the Laplace-Runge-Lenz vector of the electron is proportional to the energy in the linear Stark shift and can be accessed by exciting Stark states.

The minor changes required in the procedure for calculating the cross section associated with exchange ionization is discussed in Sec. \ref{Exchange_ionization}.

\subsection{Numerical Method} \label{numerical}
The classical equations of motion~[Eqs.~(\ref{acceleration}) and (\ref{velocity})] are solved numerically using the RK6 method with an adaptive step-size \cite{numericalrecipes}. The RK6 method does not conserve the phase space volume. Therefore, the total energy of the system which ideally should be conserved will either increase or decrease with time. For a given time-step, the larger the acceleration of the particle, the larger is the error in the velocity and position. At every instance of time, the step size is varied based on the acceleration experienced by the particles, so that the error is below an acceptable threshold~(see Sec.~\ref{convergence}). For every Monte Carlo run, the aim is to carry out these calculations until an ionization is detected or until the Rydberg atoms pass each other.

The expression in Eq. (\ref{netforce}) being a purely Coulombic potential can lead to singularities in acceleration during a trajectory. This will cause the adaptive step size algorithm to yield a time-step which can be quite small, thus resulting in runs that may not be feasible computationally. To handle this issue, we override this very small time-step with a threshold value and proceed with the RK6 method directly for the next time-step only. If this process gets repeated, this may lead to a buildup of errors in the total energy with time. If the error in the total energy exceeds 0.1$\%$ of the initial total energy, the run is classified as a failed run and not used in the cross section calculations. The number of failed runs are kept well under 3\% of the total number of Monte Carlo runs for all the scenarios discussed in this paper, except for the case of $l=0.2n$ in Fig. \ref{Frisbees_LRLscalarvaried} for which the failed runs are closer to 6\%. Note that there exists a trade-off between using a purely Coulombic potential and dealing with failed runs or resorting to a softcore Coulombic potential with no failed runs but having to accept less realistic results.
\subsection{Convergence}\label{convergence}
Here we briefly discuss the choice of various parameters defined in the previous section. We find that the probability of ionization $P_{ion}$ depends on the distance of separation between the two atoms $L_{sep}$ for $L_{sep} < 60 \cdot R_{Ryd}$.  However, for $L_{sep} > 60 \cdot R_{Ryd}$, the probability of ionization $P_{ion}$ and hence the ionization cross section does not change beyond statistical fluctuations associated with the Monte Carlo approach. Note that this factor of 60 is independent of $n$ and is obtained by numerical experimentation.

We probe the convergence of the cross section with respect to the maximum impact parameter $b_{max}$ by restricting our atoms to be exactly on the circle of radius $0.5 \cdot b_{max}$. Then, we search for a threshold value for the radius $b_{max}$ beyond which strictly no case of ionization occurs from an entire set of runs. Upon varying the radius $b_{max}$, we find that for $b_{max} > 5 \cdot R_{Ryd}$, no case of ionization is reported for the chosen $V_{CM}$~(refer.~Sec.~\ref{ionization_crosssection}). 

For a given successful run, either ionization occurs or it does not. Thus, the Monte Carlo runs form the population of a binomial distribution. For 10,000 runs, the standard deviation in the distribution is under 5\% of the mean of the distribution for the case with the lowest cross section.

\section{Applications} \label{applications}
\subsection{LRL scalar} \label{LRL_scalar}
Before we apply the procedure developed in Sec. \ref{Methods} to study the ionization cross sections, we briefly discuss a quantity which is an approximate constant of motion for large atom separations. For a given set of initial conditions, let the Laplace-Runge-Lenz vector for each electron be $\boldsymbol{A_1}$ and $\boldsymbol{A_2}$ respectively. 

The Laplace-Runge-Lenz vector is a constant of motion for a central inverse square force~\cite{goldstein_poole_safko_2011}. Given the interaction of other charges on a single electron and the motion of the nucleus, the Laplace-Runge-Lenz vector will deviate from its initial value with time. If the atoms are far apart, the Laplace-Runge-Lenz vector precesses with the components of the vector simply oscillating about a mean value. But, as the atoms approach each other and the interaction strength between the two atoms grows, the components of the Laplace-Runge-Lenz vector will start to deviate from the initial value.

We define a quantity LRL scalar $\Gamma$ as follows,
\begin{equation} \label{LRLscalar_defn}
    \Gamma = \boldsymbol{A}_1 \cdot \boldsymbol{A}_2 - 3(\hat{\boldsymbol{R}}\cdot\boldsymbol{A}_1 )(\hat{\boldsymbol{R}}\cdot\boldsymbol{A}_2 )
\end{equation}
where, $\hat{\boldsymbol{R}}$  refers to a unit vector from the nucleus of atom 2 to nucleus of atom 1. An underlying motivation for this definition arises from quantum mechanics. In a quantum mechanical system, this quantity commutes with the Hamiltonian within the n-manifold if the interaction between the two atoms is approximated as a dipole-dipole interaction~\cite{dipole_LRL_Robicheaux}. Also note that in quantum mechanics, the definition of Laplace Runge Lenz vector is modified to account for the non-commutation of  $\boldsymbol{p}$ and  $\boldsymbol{L}$ by using the symmetric form ($\boldsymbol{p} \times \boldsymbol{L}$ - $\boldsymbol{L} \times \boldsymbol{p})/2$~\cite{merzbacher1998quantum}. The LRL scalar [Eq.~(\ref{LRLscalar_defn})] is non-zero only for a pair of Rydberg atoms in elliptical orbits since for circular orbits, the Laplace-Runge-Lenz vector of both the atoms vanish. 

If two atoms which are at rest are placed relatively close to each other ($L_{sep}$ = $8 \cdot R_{Ryd}$), we find that the LRL scalar $\Gamma$ which resembles the dipole-dipole interaction energy oscillates significantly less than the individual components of the Laplace-Runge-Lenz vector. However, note that this quantity does not remain a constant when the atoms are close enough to each other such that the dipole approximation breaks down. The LRL scalar is proportional to the electric dipole-dipole interaction energy between the two Rydberg atoms for a given principal quantum number $n$ as the individual Laplace-Runge-Lenz vectors are proportional to the electric dipole moment of each atom~\cite{dipole_LRL_Robicheaux}. We examine if the initial LRL scalar of the two atoms when they are far away can be used to effectively characterize their ionization tendencies as they approach each other. We do this (refer Sec.~\ref{Section_vary_Ldir}) by analyzing the correlations between $\Gamma$ and the ionization cross section of the Rydberg atoms.

It should be noted that $\Gamma$ changes slightly between Monte Carlo runs. While $\boldsymbol{A_1}$ and $\boldsymbol{A_2}$ remain constant for a given set of runs, the positions of each atom and hence $\hat{\boldsymbol{R}}$ changes slightly with every run. This is because the randomizing of the initial conditions for the Monte Carlo approach, involves a change in the impact parameter (not $b_{max}$) and a small change in the $L_{sep}$~(refer Sec.~\ref{Methods}).

\subsection{Ionization Cross section} \label{ionization_crosssection}
Here, we investigate how the ionization cross section depends on the magnitude and the direction of orbital angular momentum of the electron and the direction of the Laplace-Runge-Lenz vector. In our calculations, $V_{CM} = 10^{-4}$~a.u. which corresponds to the rms speed at a temperature of $\sim$150 K. The chosen $V_{CM}$ is similar to the values observed in the experiment by Fields et al.~\cite{dunning_destructioncollision}. The results of our calculation were relatively unchanged for a range of velocities between 7 x $10^{-5}$ to 2 x $10^{-4}$ a.u which corresponds to the kinetic energies associated with a temperature range of approximately 80K to 650K. A more detailed analysis of the dependence of the ionization cross section on the relative velocity of the Rydberg atoms can be found in Ref.~\cite{olson1979ionization_vcm}. 

While the trajectories of the particles including the relative motion of the two nuclei are not assumed to be straight lines, the calculations indicate little deviation from the straight-line trajectory for the two nuclei. This is partly because of their large mass. In order for the collision energy of nuclei to become comparable to their interaction energy and their trajectories to deviate from a straight line, the thermal velocities of the atoms should be smaller by about two orders of magnitude from the chosen value ($V_{CM} = 10^{-4}$~a.u.).

%

\begin{figure}
\resizebox{80mm}{!}{\includegraphics{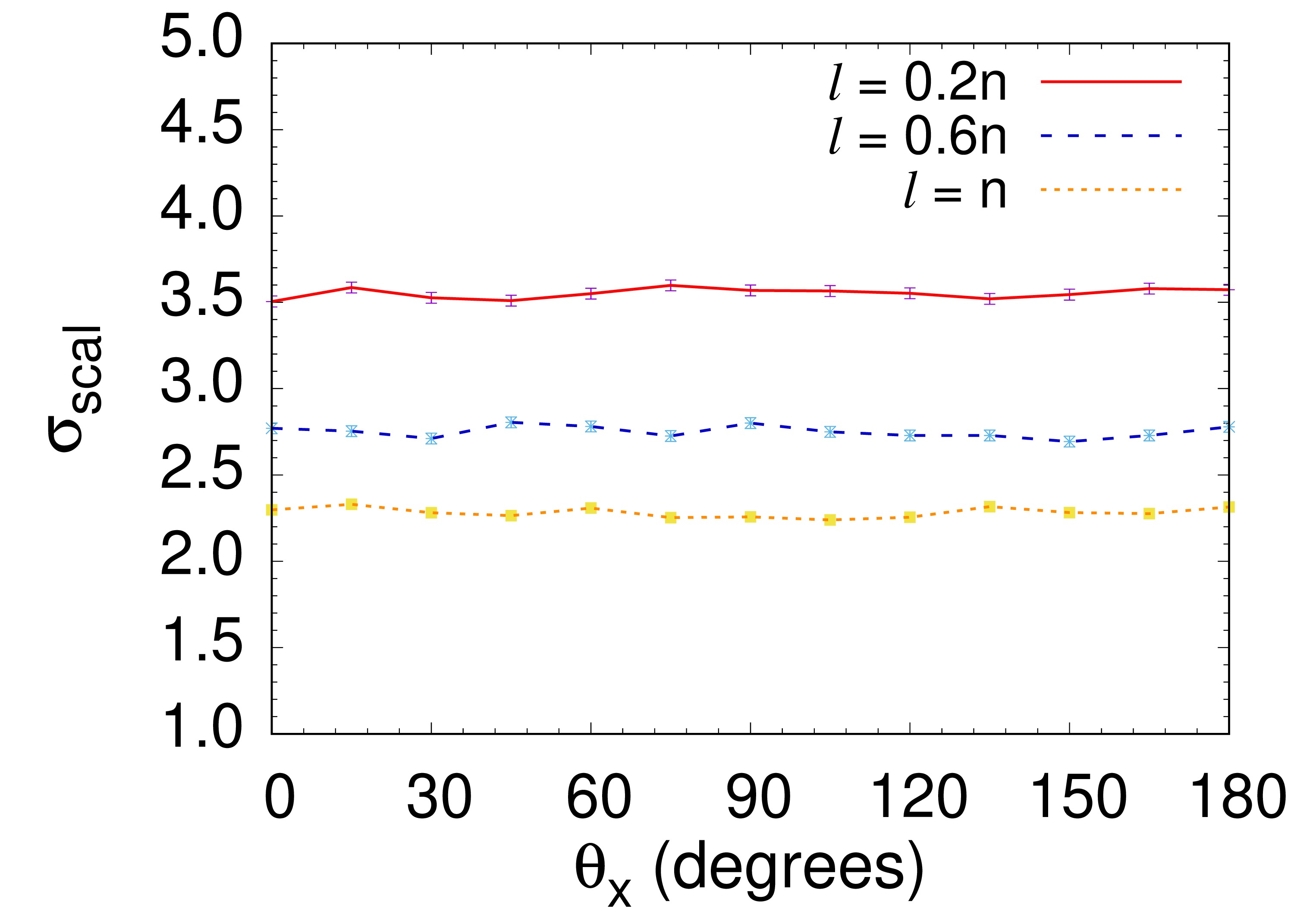}}
\caption{\label{Frisbees_Ldirectionvaried}
The figure shows a plot of scaled ionization cross section as a function of rotation angle about the x-axis, for the initial orientation of type-Frisbees. 
Each point is a result of 10,000 Monte Carlo runs. The error bars indicate the standard deviation in the cross section. Here, rotation about the x-axis changes the direction of angular momentum but preserves the direction of the Laplace-Runge-Lenz vectors. The points with the same initial angular momentum $l$, have been connected to serve as a visual cue. For a given $l$, the ionization cross section does not change with $\theta_x$. This is expected from rotational symmetry.
}
\end{figure}

\begin{figure}
\resizebox{80mm}{!}{\includegraphics{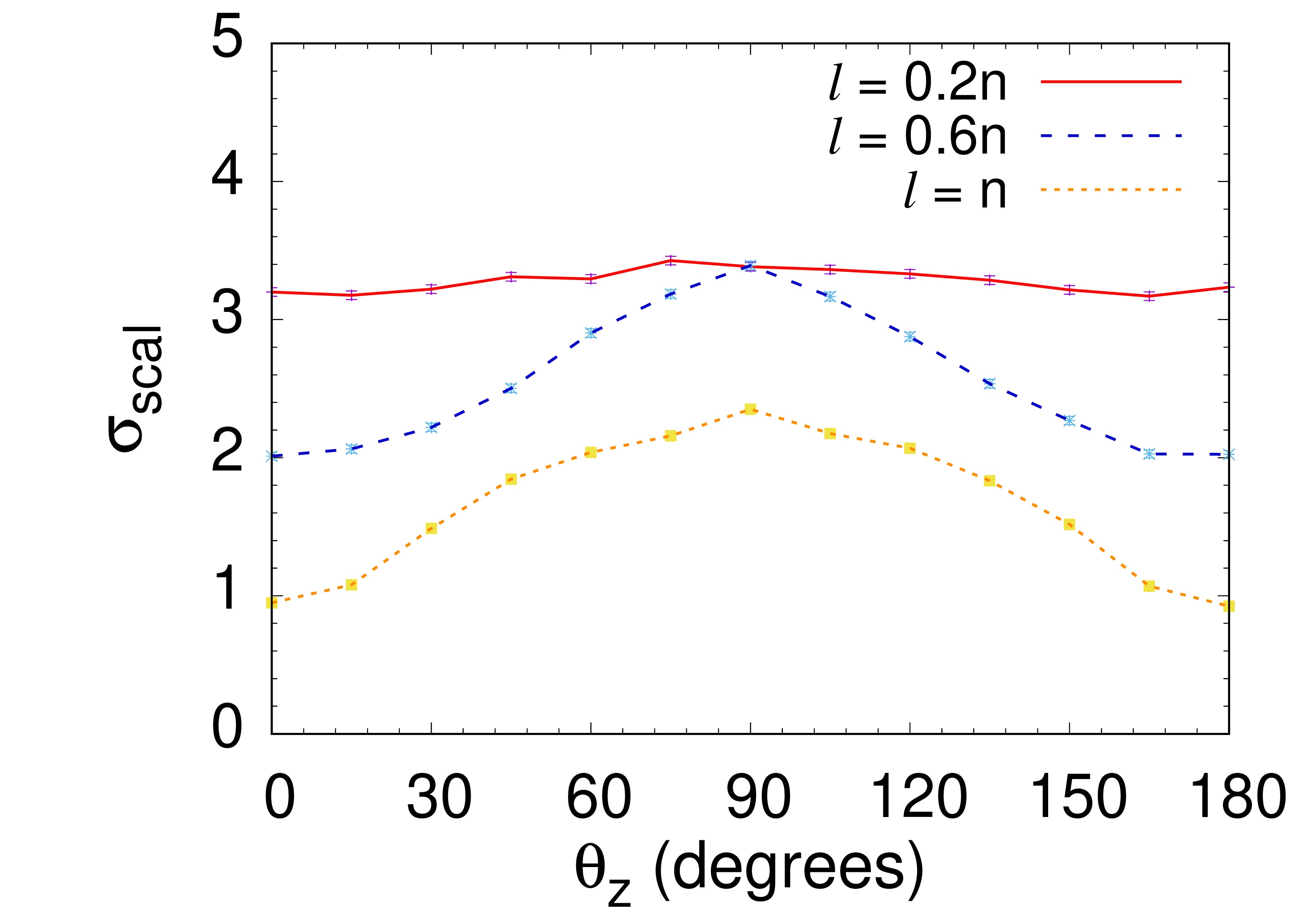}}
\caption{\label{Pancakes_Ldirectionvaried}
The figure shows a plot of scaled ionization cross section as a function of rotation angle about the z-axis, for the initial orientation of type-Cymbals. 
Each point is a result of 10,000 Monte Carlo runs. The error bars indicate the standard deviation in the cross section. Here, rotation about the z-axis changes the direction of angular momentum but preserves the direction of the Laplace-Runge-Lenz vectors. The points with the same initial angular momentum $l$, have been connected to serve as a visual cue. 
}
\end{figure}

\subsubsection{Varying the direction of the angular momentum} \label{Section_vary_Ldir}
We explore the effect of varying the direction of the angular momentum of the electrons $\boldsymbol{L_1}$ and $\boldsymbol{L_2}$ on the ionization cross section. Here, we preserve the direction of the Laplace-Runge-Lenz vectors $\boldsymbol{A_1}$ = $\boldsymbol{A_2}$ while we change the direction of the angular momentum vectors, $\boldsymbol{L_1}$ = $\boldsymbol{L_2}$. 
Based on the discussions in Sec. \ref{Methods}, we consider two possible initial orientations, Frisbees (see Fig. \ref{Diagram_fris}) and Cymbals (see Fig. \ref{Diagram_pan}). 

For the initial orientation of type-Frisbees, the direction of angular momentum
can be changed by rotating each atom about their respective Laplace-Runge-Lenz vector and hence the x-axis. This ensures that $\boldsymbol{A_1}$ and $\boldsymbol{A_2}$ are invariant. Therefore, the quantity $\Gamma$ [Eq. (\ref{LRLscalar_defn})] remains an invariant during this rotation.

For the initial orientation of type-Cymbals, the direction of the angular momentum can be changed by rotation of the atoms about z-axis. This ensures that $\boldsymbol{A_1}$ and $\boldsymbol{A_2}$ remain the same. Again, $\Gamma$ remains invariant during the rotation for a given Monte Carlo run  but varies slightly between runs for reasons discussed in Sec.~\ref{LRL_scalar}.

It is to be noted that for type-Frisbees, for the entire range of the rotation angle, the initial orientation of the two atoms remains as type-Frisbees. But, for type-Cymbals, during the rotation about z-axis, the orientation changes from type-Cymbals to type-Frisbees at an angle of 90$\degree$.

For each orientation of the angular momentum, the ionization cross section is calculated using the procedure developed in Sec.~\ref{Methods}. Given the direction of the Laplace-Runge-Lenz vector, a rotation angle range of 0-180 degrees exhaustively covers all possible relative orientations between the two atoms for the angular momentum.  
These calculations are performed for different magnitudes of the angular momentum (see Figs. \ref{Frisbees_Ldirectionvaried} and \ref{Pancakes_Ldirectionvaried} ). 

For type-Frisbees~(Fig.~\ref{Frisbees_Ldirectionvaried}), we find that the ionization cross section is independent of the rotation angle about x-axis for a given angular momentum magnitude, $l$. This is expected from the rotational symmetry as the angle between the angular momentum vector and $\hat{\boldsymbol{R}}$ remain the same throughout the rotation given the direction of Laplace Runge Lenz vector~(Fig.~\ref{Diagram_fris}). Upon increasing $l$, we find that the cross section decreases. This indicates that an electron in a circular orbit is more difficult to ionize than an electron in a highly elliptical orbit. This becomes clear if we examine the outer turning point radius($r_{out}$) for elliptical orbits which is given by~\cite{turningpoint_equation},  
\begin{equation} \label{turningpoint_equation}
    \frac{r_{out}}{R_{Ryd}} = 1 + \sqrt{1 - \bigg(\frac{l}{n}\bigg)^2} 
\end{equation}
So, for an orbit with $l=0.2n$ the outer turning point distance from the nucleus is approximately twice as large when compared to a circular orbit ($l=n$) of the same energy ($n$). Thus making it likely that the electron will venture into regions of stronger field from the other atom. 

\begin{figure}
\resizebox{80mm}{!}{\includegraphics{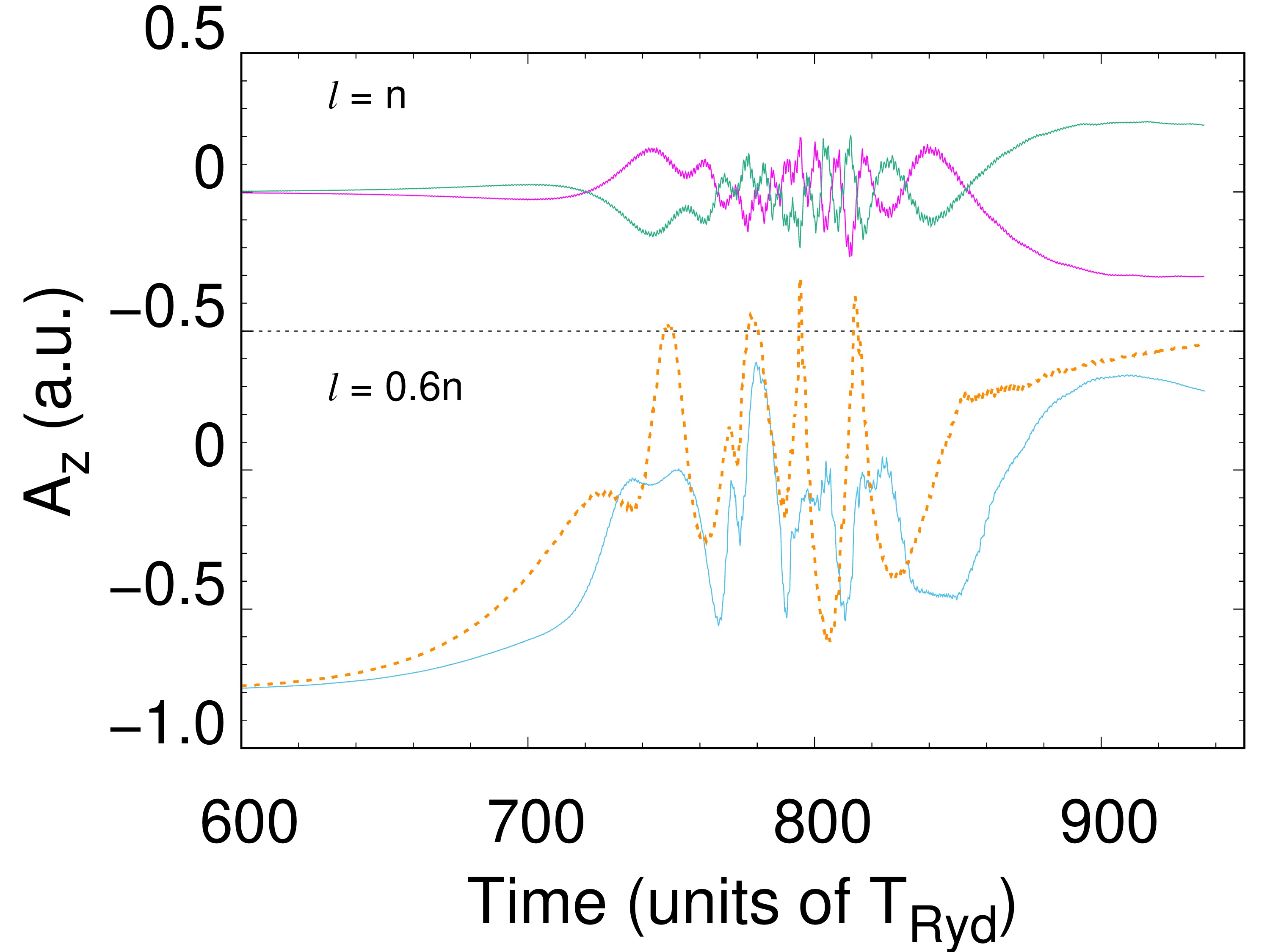}}
\caption{\label{single_trajectory_pan}
The figure shows a plot of the z-component of the Laplace-Runge-Lenz vector of each electron as a function of time for a typical non-ionizing run, for the initial orientation of type-Cymbals ($\theta_z = 0\degree$ in Fig.~\ref{Pancakes_Ldirectionvaried}). The bottom pair of curves are for the case of $l=0.6n$, with the blue and the orange (dotted) line representing electron 1 and electron 2 respectively. The top pair of curves are for the case of $l=n$ (circular orbits) with the magenta (starts at the bottom) and green line (starts at the top) representing electron 1 and electron 2 respectively. Note that, these two lines mirror each other. The black dotted line between the top pair and the bottom pair separates the y-axis of the two plots. From the bottom pair of curves, it is evident that the elliptical case lacks the stabilizing oscillations seen in the circular case. 
}
\end{figure}

For type-Cymbals~(Fig.~\ref{Pancakes_Ldirectionvaried}), for small magnitudes of angular momentum we find that the cross section does not appear to change with $\theta_z$. As the $l$ value increases, the cross section peaks at a rotation angle of $\theta_z$ = 90$^\circ$. The physics behind this can be understood in the following manner: Consider the case of two circular ($l=n$) Rydberg atoms of type-Cymbals (Fig.~\ref{Diagram_pan}) approaching each other, the electrons in these atoms tend to push each other to the opposite extremes of their orbits. This results in the atoms becoming more elliptical as they approach each other or in other terms, $\boldsymbol{A_1}$ and $\boldsymbol{A_2}$ start building up from zero in opposite directions.
This can help in minimizing the interaction energy, thus lowering the ionization cross section effectively. This is evident from the calculations for a single collisional run~(Fig.~\ref{single_trajectory_pan}) that does not result in an ionization. These calculations shows that as the two circular Rydberg atoms of type-Cymbals approach each other their respective Laplace-Runge-Lenz vectors start building up in opposite directions and exhibits oscillatory behaviour.

This effect is less likely to occur as you make the orbits elliptical~(decrease $l$). The reason being if the Laplace-Runge-Lenz vectors were already non-zero and equal to each other in the beginning, as they approach each other it becomes more difficult to get them to orient in opposite directions given the symmetric nature of the two atoms. This is also evident from the calculations for a single collisional run~(Fig.~\ref{single_trajectory_pan}) that shows the difficulty in achieving similar stable oscillatory behaviour found in the circular case. The z-component of the Laplace-Runge-Lenz vector  has been chosen to illustrate how the non-zero value in the elliptical case hinders this oscillatory behaviour. Note that the other components of the Laplace-Runge-Lenz vector exhibit similar behaviour for the circular case. In the elliptical case, there is a lack of this type of oscillatory behaviour in all of the three components of the Laplace-Runge-Lenz vector.

For the case of type-Frisbees~(Fig.~\ref{Diagram_fris}), this effect will not manifest. The nature of the orientation offers significantly lesser amount of time for $\boldsymbol{A_1}$ and $\boldsymbol{A_2}$ to gradually buildup in opposite directions. 
Single collisional run calculations for Rydberg collisions of type-Frisbees consistently show the absence of the anti-parallel oscillatory behaviour of $\boldsymbol{A_1}$ and $\boldsymbol{A_2}$ found in the case of type-Cymbals~(Fig.~\ref{single_trajectory_pan}). This effect lends some perspective into why ionization cross section from Fig.~\ref{Pancakes_Ldirectionvaried} for type-Cymbals is lower by a factor of $\sim2-3$ than that for type-Frisbees.

\begin{figure}
\resizebox{80mm}{!}{\includegraphics{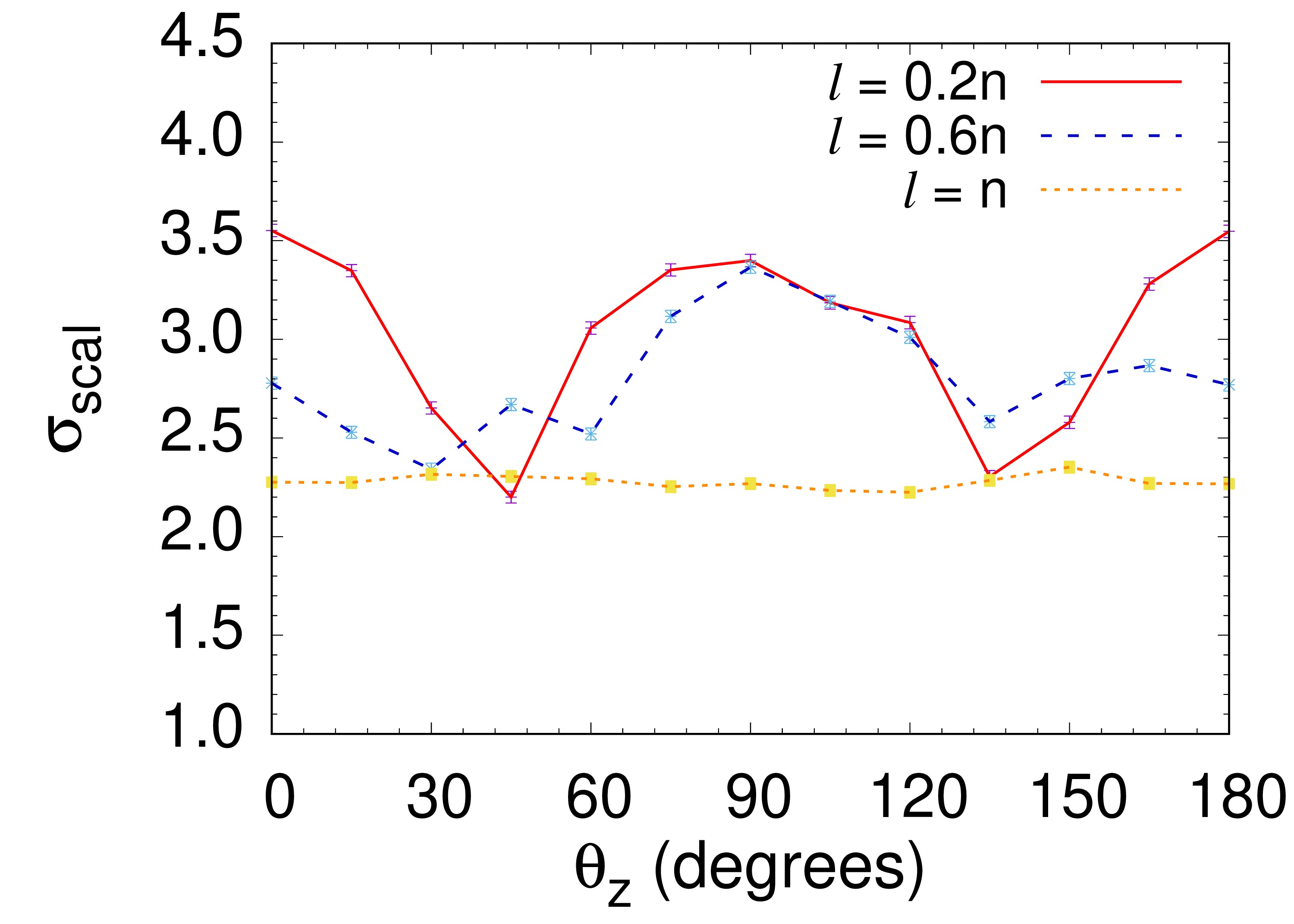}}
\caption{\label{Frisbees_LRLscalarvaried}
The figure shows a plot of scaled ionization cross section as a function of rotation angle about the z-axis, for the initial orientation of type-Frisbees. 
Each point is a result of 10,000 Monte Carlo runs. The error bars indicate the standard deviation in the cross section. Here, the rotation about the z-axis changes the direction of Laplace-Runge-Lenz vectors but preserves the direction of angular momentum. This changes the value of the LRL scalar $\Gamma$. The points with the same initial angular momentum $l$, have been connected to serve as a visual cue.
}
\end{figure}

Although a rotation angle of $\theta_z$ = 90$^\circ$ corresponds to a orientation of type-Frisbees, the results from Fig.~\ref{Pancakes_Ldirectionvaried} should not be compared to Fig.~\ref{Frisbees_Ldirectionvaried}. The reason for this is because, in  Fig.~\ref{Frisbees_Ldirectionvaried}, the Laplace-Runge-Lenz vector is chosen to be along the x-axis but the the Laplace-Runge-Lenz vector in Fig. \ref{Pancakes_Ldirectionvaried} for $\theta_z = 90^\circ$ is along the negative z-axis. However for the special case of $l=n$, the Laplace-Runge-Lenz vector is zero and a comparison of the two figures at $\theta_z = 90^\circ$ reveals a good agreement. In the next section, we discuss cases where the Laplace-Runge-Lenz vector is varied. One may then compare the results of Fig.~\ref{Pancakes_Ldirectionvaried} with Fig.~\ref{Frisbees_LRLscalarvaried} for the value of $\theta_z$ = 90$^\circ$ in both figures. This comparison shows a good agreement.


\begin{figure}
\resizebox{80mm}{!}{\includegraphics{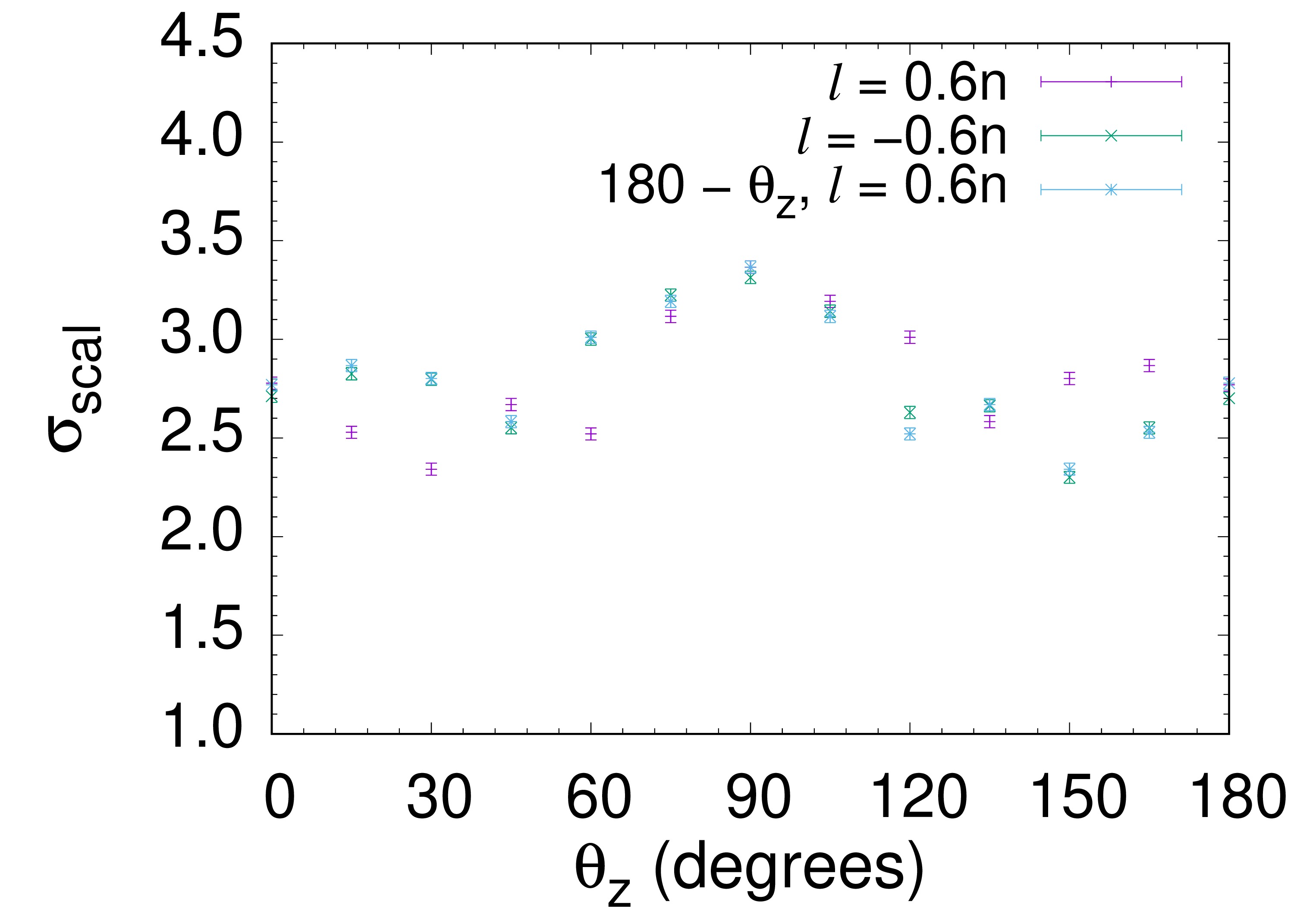}}
\caption{\label{l_negative_comparison}
The plot shows an analogous calculation to Fig. \ref{Frisbees_LRLscalarvaried}, for the case where the direction of angular momentum has been inverted ($l=-0.6n$). For comparison, the case of $l=0.6n$ is plotted after reflection about 90$\degree$. In simpler terms, for the case of $l=0.6n$, an angle of 30$\degree$ in Fig. \ref{Frisbees_LRLscalarvaried} corresponds to an angle of 150 $\degree$ in the above figure. A comparison of the two cases $l=-0.6n$ and $180 - \theta_{z} , l=0.6n$ reveals very good agreement. This clearly indicates that the direction of angular momentum plays a role in the asymmetry.
}
\end{figure}

\begin{figure}
\resizebox{80mm}{!}{\includegraphics{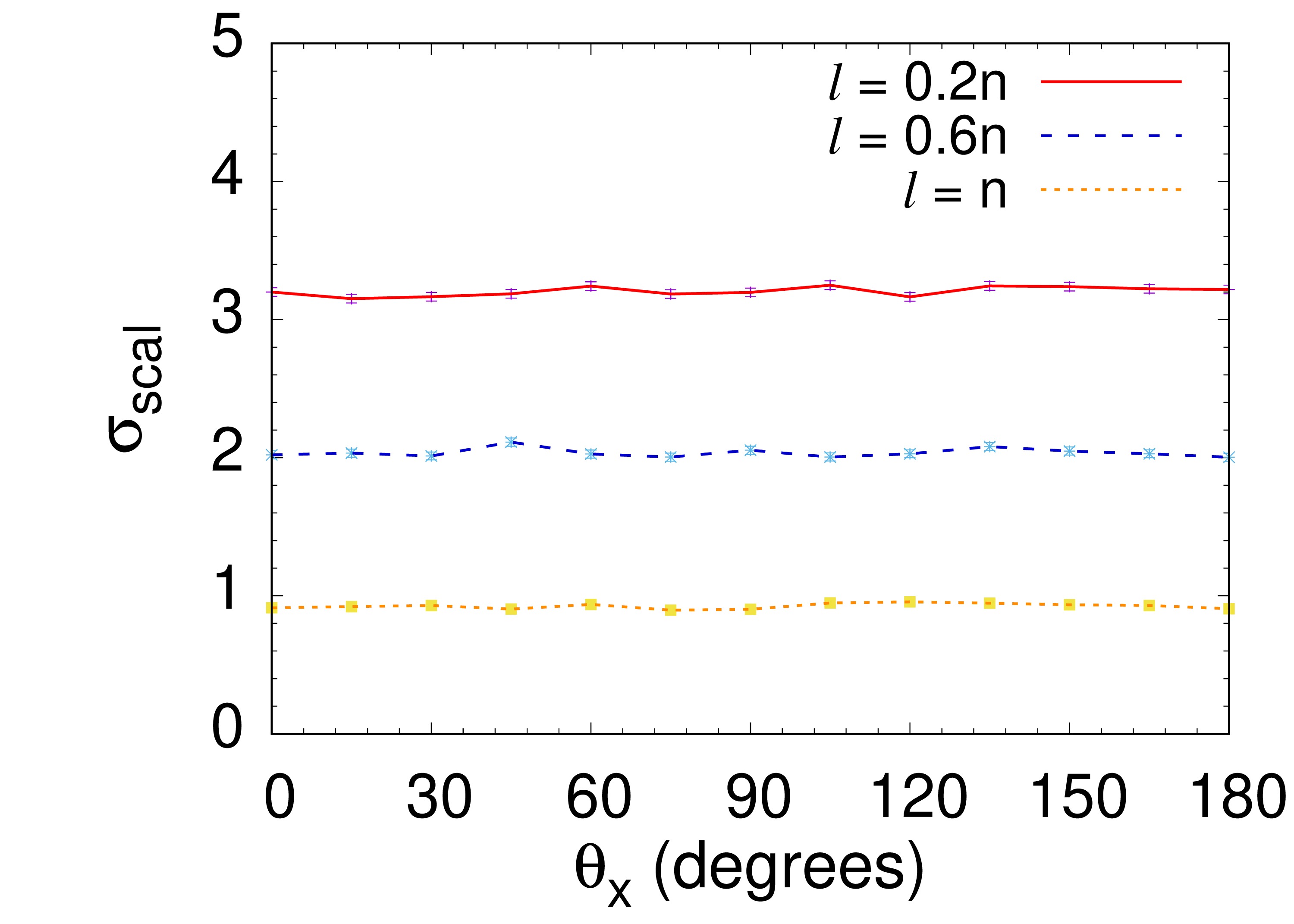}}
\caption{\label{pancakes_LRLscalar}
The figure shows a plot of scaled ionization cross section as a function of rotation angle about the x-axis, for the initial orientation of type-Cymbals. 
Each point is a result of 10,000 Monte Carlo runs. The error bars indicate the standard deviation in the cross section. Here, the rotation about the x-axis changes the direction of Laplace-Runge-Lenz vectors but preserves the direction of angular momentum. It is to be noted that this rotation does not appreciably change the LRL scalar $\Gamma$ because for this configuration $\hat{\boldsymbol{R}}$ remains largely perpendicular to $\boldsymbol{A_1}$ and $\boldsymbol{A_2}$. The points with the same initial angular momentum $l$, have been connected to serve as a visual cue.
}
\end{figure}

\subsubsection{Varying the LRL scalar} \label{Section_vary_LRLscalar}
We explore the dependence of the ionization cross section on $\Gamma$. We vary $\Gamma$ by changing the direction of Laplace-Runge-Lenz vectors $\boldsymbol{A_1}$ and $\boldsymbol{A_2}$ but preserving the direction of the angular momentums $\boldsymbol{L_1}$ and $\boldsymbol{L_2}$. We do this  by rotating each atom about their respective angular momentum vectors.
This is equivalent to rotation about z-axis and x-axis for the initial orientation of type-Frisbees~(Fig.~\ref{Diagram_fris}) and type-Cymbals~(Fig.~\ref{Diagram_pan}) respectively. Note that this rotation only changes the LRL scalar for type-Frisbees and not type-Cymbals as the angle between the Laplace-Runge-Lenz vector of each atom and $\hat{\boldsymbol{R}}$ remains relatively unchanged for type-Cymbals.

We calculate the ionization cross section, average initial $\Gamma$ and the standard deviation in initial $\Gamma$ for a given angle of rotation, from a set of Monte Carlo runs. Note that the average $\Gamma$ and the standard deviation in $\Gamma$ change with rotation angle and $l$. Again, we repeat these calculations for different angles of rotation and different magnitudes of the angular momentum $|\boldsymbol{L_1}|$ and $|\boldsymbol{L_2}|$.

Consider the initial configuration of type-Frisbees (see Fig. \ref{Frisbees_LRLscalarvaried}), 
for $l= 0.2n$, the average value of the LRL scalar increases from $-1.910\pm0.005$ a.u. at $0\degree$ to a maximum value of $0.955\pm0.005$ a.u. at $90\degree$, only to revert back to the value of $-1.910\pm0.005$ a.u. at $180\degree$. There appears to be a positive correlation between the ionization cross section and the modulus of the LRL scalar, $\Gamma$. The average LRL scalar exhibits similar behaviour for other $l$ values, except that the average LRL scalar decreases with increase in $l$. 

This observed positive correlation should be expected because the quantity LRL scalar $\Gamma$ is proportional to the electric dipole-dipole interaction energy~(refer Sec.~\ref{LRL_scalar}). A large absolute initial value for $\Gamma$ implies a large interaction energy between the two atoms when they are far apart. If $\Gamma$ is an adiabatic invariant until the atoms get close to each other (see Sec.~\ref{LRL_scalar}), then a large initial interaction energy implies a large interaction energy when they are relatively close and thus consequentially one might expect a large ionization cross section.

An interesting observation from Fig.~\ref{Frisbees_LRLscalarvaried} is that there exists an asymmetry about $\theta_{z} = 90\degree$ for intermediate values of angular momentum. This is easily noticeable for $l = 0.6n$. Intuitively, one might expect a symmetry because of how the relative orientations of the Laplace-Runge-Lenz vectors between the two atoms remain unchanged for the two rotation angles $\theta_{z}$ and $180 - \theta_{z}$.  A supporting argument for this expectation is the fact that the LRL scalar $\Gamma$ is found to be symmetric about 90$\degree$. However, a careful analysis reveals that this symmetry is broken by the direction of the angular momentum. An identical calculation to the one in Fig. \ref{Frisbees_LRLscalarvaried} for $l=0.6n$ but with the direction of angular momentum reversed verifies this (See Fig. \ref{l_negative_comparison}). Interestingly, a similar asymmetry was found to exist in Ion-Rydberg collisions~\cite{Asymmetry_KBmacadam}. 

For the initial configuration of type-Cymbals, the rotation about the x-axis changes the direction of the Laplace-Runge-Lenz vectors. But, it does not change the value of LRL scalar $\Gamma$. The results from the calculation (see Fig. \ref{pancakes_LRLscalar}) for the ionization cross section indicate that for a given magnitude of angular momentum $l$, the cross section remains a constant with this rotation. This is expected because of symmetry as the rotation still preserves the relative orientation of the two atoms and is equivalent to merely observing the collision from a different viewing angle. This result is consistent with the earlier observation that the ionization cross section is positively correlated with the modulus of the LRL scalar $\Gamma$.


 

\subsection{Exchange Ionization} \label{Exchange_ionization}
Here we focus exclusively on exchange ionization which is distinct from simple ionization. By simple ionization, we refer to those collisions in which the atom that loses the electron finally ends up being ionized after they pass each other. This is different from exchange ionization where, atom 1(2) initially loses its electron but captures the electron from atom 2(1) and atom 2(1) ends up being ionized. Atom 1(2) however departs with electron 2(1). Note that the procedure described in Sec.~\ref{Methods} counts all possible ionizations including exchange ionizations. 

Exchange ionizations can be exclusively counted by running every Monte Carlo run until the Rydberg atoms pass each other and then classifying the ionization that may occur as an exchange ionization if the distance between the nucleus of atom 1(2) and electron 2(1) is within 2$R_{Ryd}$. This restriction has been chosen by measuring the final distances between the exchanged electron and the nucleus across multiple runs and ensuring that all exchanges are included. 
We can calculate the exchange ionization cross section from the corresponding probability by using the same equation used to determine the total ionization cross section~[Eq. (\ref{cross_sec_defn})]. 

These calculations reveal the following: for the initial orientation of type-Frisbees with $l = n$, we get a scaled exchange ionization cross section of $1.16 \pm 0.02 $. For reference, the scaled total ionization cross section for the same case is $2.23 \pm 0.03 $. This indicates that exchange ionization contributes to as much as 50\% of the total ionizations. The calculations show that the exchange ionization also increases with decrease in the magnitude of the orbital angular momentum of the electrons analogous to the total ionization cross section. This can be interpreted by the same argument why circular orbits are more stable (see Sec.~\ref{Section_vary_Ldir}). As the orbits become more elliptical their outer turning point distances increases and consequentially they can be more easily captured by the other atom and also are more likely to be the reason for the ionization of the other electron. This is supported by the fact that single collisional run calculations indicate that the capture and the ionization occur relatively at the same time. 

While the identical nature of electrons makes it difficult to study exchange ionization experimentally, one way it can be studied is by creating atom 1 with the valence electron in a state of spin-up and atom 2 with the valence electron in a state of spin-down. This way we have effectively labelled the electrons and can in principle track their final states. 
\\

\section{Conclusion and Summary}
We defined a quantity called LRL scalar $\Gamma$ and discussed how the fluctuations in this quantity were much lower than the fluctuations in the individual Laplace-Runge-Lenz vectors $\boldsymbol{A_1}$ and $\boldsymbol{A_2}$ of the electrons. We studied the dependence of the ionization cross section on the direction and the magnitude of the orbital angular momentum of the electron and the direction of the Laplace-Runge-Lenz vectors of the electrons. 

These calculations revealed the following: First, the ionization cross sections exhibited positive correlations with the modulus of the LRL scalar [Eq.~(\ref{LRLscalar_defn})]. The underlying reason being that the electric dipole-dipole interaction energy is proportional to the LRL scalar at large atom separations and therefore the ionization cross section would increase with the interaction. Second, the Rydberg atoms with highly elliptical orbits (small $l$) had higher ionization cross sections relative to Rydberg atoms with circular orbits ($l = n$) implying that circular orbits were significantly more stable to collisional ionization. Third, as the magnitude of the angular momentum is increased ($l \rightarrow n$), the initial configuration of type-Cymbals for the atoms exhibited significantly lower ionization cross sections (lower by a factor of $\sim 2-3$ for $l = n$) than that of type-Frisbees. This was intrepreted in terms of the tendency of the Laplace-Runge-Lenz vectors to become anti-parallel as the atoms approach each other.

Finally, exchange ionization was studied and its dependence on the magnitude of angular momentum was found to be similar to that of the total ionization cross section. The calculations indicated that the exchange ionizations contributed to about 50\% of the total ionizations. 

These results indicate several ways in which the stability of Rydberg atoms against collisional ionization can be significantly improved. The lowest ionization cross section is achieved for the case of circular orbits and when the relative orientation of the two atoms is of type-Cymbals.

\section{Acknowledgements}
This work was supported by the U.S. Department of Energy, Office of
Science, Basic Energy Sciences, under Award No. DE-SC0012193. A.V. is grateful to S.Pandey for discussions on the nomenclature of the two initial orientations. A.V. thanks S. Vaidya for the sanity-check discussions. F.R. benefited from discussions with and early modeling by S.F. Behmer.

\bibliography{References.bib}

\end{document}